\documentclass{ttp_DSL2006}

\usepackage[dvipsone]{graphicx}
\usepackage[intlimits]{amsmath}
\usepackage{amssymb}
\usepackage{exscale}
\usepackage[colorlinks,urlcolor=blue]{hyperref}

\begin{document}

\title{ Functional-integral approach to the investigation of the spin-spiral magnetic order and phase separation} 

\author{A.K.~Arzhnikov \inst{1} \textbf{,} A.G.~Groshev \inst{2} }

\institute{Physical-Technical Institute, Ural Branch of RAS, Kirov str. 132, Izhevsk 426001, Russia}
\maketitle

\vspace{-3mm}
\sffamily
\begin{center}
$^{1}$arzhnikov@otf.pti.udm.ru $^{2}$Groshev$_{-}$a.g@mail.ru
\end{center}

\vspace{2mm} 
\hspace{-7.7mm} 

\vspace{-2mm} \hspace{-7.7mm}
\rmfamily
%\hspace{-7.7mm} 
%\noindent \textbf {\textit {For~the~rest~of~the~paper,~please~use~Times~Roman~(Times~New~Roman)~12}} \\ 
\noindent \textbf{Abstract.} 
We investigate a two--dimensional single-band Hubbard model with a nearest--neighbor 
hopping. We treat a commensurate collinear order as well as incommensurate spiral 
magnetic phases at a finite temperature using a Hubbard--Stratonovich transformation 
with a two--field representation and solve this problem in a static approximation. 
We argue that temperature dramatically influence the collinear and spiral magnetic 
phases, phase separation in the vicinity of half--filling. The results imply a possible 
interpretation of unusual behavior of magnetic properties of single--layer cuprates. 

\vspace{-2mm}
\section{Introduction}
\vspace {-3mm}

\hspace{4.5mm} 

Investigation of two--dimensional (2D) electronic systems attracts substantial interest, 
which has been stimulated by the discovery of high-temperature superconducting cuprates. 
It is generally accepted that superconducting and magnetic propeties of cuprates are closely 
related, but magnetic properties are formed by $Cu$--layers and interlayer coupling is weak. 
At half--filling, the cuprates are antiferromagnetically ordered, evolution of their magnetic 
properties with doping and temperature is an interesting challenge. For example, neutron 
scattering in $La_{2-x}Sr_{x}CuO_{4}$ reveals coexistence of both commensurate and 
incommensurate magnetic structures in the vicinity of half--filling at low temperature, 
that pass to antiferromagnetic with rise of temperature 
\cite{Yamada_1998,Matsuda_2000,Fujita_2002,Matsuda_2002}.
Recently the authors \cite{Igoshev_2010} considered the ground--state magnetic phase diagram 
of the two--dimensional single-band Hubbard model with nearest and next--nearest--neighbor 
hopping in terms of electronic density and interaction. They treated commensurate 
ferromagnetic and antiferromagnetic as well as incommensurate spiral magnetic phases using 
the mean field (MF) approximation. First--order magnetic transitions with changing 
chemical potential, resulting in a phase separation (PS) in terms of density, was found 
between ferromagnetic, antiferromagnetic, and spiral magnetic phases. 

Here phase diagram is investigated depending on temperature. Our calculations are based 
on a two--dimensional single--band Hubbard model. We use a Hubbard--Stratonovich 
transformation with a two--field representation and solve this problem in a static 
approximation. It allows us, on the one hand, to obtain a solution that coincides with 
the MF approximation at zero temperature and, on the other hand, to investigate the temperature 
behavior of PS and phase transformation taking into account thermodynamic transverse 
spin fluctuations.

First of all, we want to emphasize the peculiarity of the 2D magnetic systems. The series 
of works proved rather directly impossibility of existence of a spontaneous magnetic order and 
phase separation in a two-dimensional Hubbard model \cite{Ghosh_1971,Su_1996,Laad_2008} (at finite 
temperature the results \cite{Ghosh_1971,Su_1996} coincide with  Mermin's theorem). At the same 
time, there are a lot of papers using the 2D Hubbard model for describing cuprates (see, for example, 
\cite{Langmann_2007,Macridin_2006,Chang_2008,Maier_2005} and references in them). Usually, it is 
believed that weak interlayer coupling stabilizes a magnetic order, which allows us to consider 
fluctuations around this broken-symmetry state even at finite temperature. Therefore, one can 
choose approximations which depress the 2D divergence of magnetic fluctuations (for example, a mean 
field approximation, a dynamic mean field approximation and so on). In addition, phase separation 
and inhomogeneous magnetic state in the 2D Hubbard model have been obtained recently by calculation 
methods using an extension of the dynamic mean-field theory and advances in a quantum Monte Carlo 
techniques \cite{Langmann_2007,Macridin_2006,Chang_2008}. We want to emphasize that exact results 
of \cite{Ghosh_1971,Su_1996,Laad_2008} were received for homogeneous states; therefore, for inhomogeneous 
case the long wave fluctuations are cut off for the mean length of homogeneity. It reminds  the 
stability of a 2D graphene layer where «ripples» appear due to thermal and quantum fluctuations and 
this stabilizes the crystal \cite{Fasolino_2007}. Though the problems of a 2D-crystal stability and 
of an inhomogeneous magnetic order in the 2D Hubbard model have not been solved yet, the results of 
the calculation methods give us an additional basis for using approximations which stabilize states 
with a broken symmetry. In this paper we neglect magnetic longitudinal fluctuations and thereby depress 
the 2D divergence.

\section{Method}
\vspace {-3mm}

We consider the Hamiltonian of the Hubbard model on the square lattice
${\hat H}={\hat H}_{0}+{\hat H}_{int},\,$
\begin{equation}
\label{eq:hamiltonian}
{\hat H}_{0}=\sum_{i,j,s}t_{ij}\hat{c}_{i,s}^{+}\hat{c}_{j,s}^{},\,\quad
{\hat H}_{int}=U\sum_{j}\hat{c}_{j,\uparrow}^{+}\hat{c}_{j,\uparrow}^{}
\hat{c}_{j,\downarrow}^{+}\hat{c}_{j,\downarrow}^{},\,
\end{equation}
where $j$ is a number of site, $s$ is a spin projection, $t_{ij}=-t$ for the 
nearest--neighbor site and $0$ if not, $\hat{c}_{j,s}^{} (\hat{c}_{j,s}^{+})$ is a creation 
(annihilation) electronic operator on the site $j$ and $U$ is the electronic (Hubbard) 
on--site interaction. 
We use a two-field formalism  in the  Hubbard-Stratonovich transformation, therefore  more convenient 
representation of the on-site interaction in form 
\begin{equation}
\label{eq:h_int}
{\hat H}_{int}=U\sum_{j}\hat{n}_{j,\uparrow}^{}\hat{n}_{j,\downarrow}^{}=
U\sum_{j}\left[\hat{n}_{j}^{2}/4-(\vec{\hat S}_{j}\cdot\vec{e}_{j})^{2}\right],\,
\end{equation}
where $\vec{e}_{j}$ is the (arbitrarily chosen) $j$--dependent unit vector and we introduce 
the site density $\hat{n}_{j,s}=\hat{c}_{j,s}^{+}\hat{c}_{j,s}^{}$, 
$\hat{n}_{j}=\hat{n}_{j,\uparrow}+\hat{n}_{j,\downarrow}$ and the site magnetization operator 
$\vec{{\hat S}}_{j}=1/2\sum_{s,s'}\hat{c}_{j,s}^{+}\vec{\sigma}_{s,s'}\hat{c}_{j,s}^{}$,  
($\vec{\sigma}_{s,s'}$ are the elements of vector 
$\vec{\sigma}=(\sigma_{x}^{},\,\sigma_{y}^{},\,\sigma_{z}^{})$ formed by the  Pauli spin matrices). 

Thermodynamical properties are defined by partition function of the grand canonical ensemble:
\begin{eqnarray}
\label{eq:Z}
Z=Tr\left[\exp\left\{-\beta({\hat H}-\mu{\hat n}\right\} \right]=
Tr\left[ \exp\left\{-\beta({\hat H}_{0}-\mu{\hat n})\right\}T_{\tau}
\exp\left\{-\int_{0}^{\beta}{\hat H}_{int}(\tau)d\tau\right\}\right].\,
\end{eqnarray}
Here $\mu$ is a chemical potential, $T_{\tau}$ is a time ordering operator, $Tr$ is summation over 
a complete set of states,
${\hat H}_{int}(\tau)=\exp{(-\tau {\hat H}_{0})}{\hat H}_{int}\exp{(\tau {\hat H}_{0})}$ is an operator 
in the interaction representation, $\beta =1/k_{B}T$, $T$ is temperature. Partition function can be 
rewrite with the help of a Hubbard--Stratonovich transformation 
\begin{equation}
\label{eq:Z'}
Z=\int\prod_{i}D{\vec v}_{j}(\tau )D\zeta_{j}(\tau )Z^{}[{\vec v},\zeta],
\end{equation}
\vspace {-5mm}
\begin{eqnarray}
Z[{\vec v},\zeta]=\exp{\left\{-\beta F({\vec v},\zeta )\right\}}=
\exp{\left\{-U\sum_{j}\int_{0}^{\beta}d\tau[{\vec v}_{j}^{2}(\tau)+\zeta_{j}^{2}(\tau)]
\right\}}Z^{*}[{\vec v},\zeta],\,
\nonumber\\
Z^{*}[{\vec v},\zeta]=Tr\left[\exp{\left\{-\beta({\hat H}_{0}-\mu{\hat n})\right\}}T_{\tau}
\exp{\left\{-\sum_{j}\int_{0}^{\beta}d\tau {\hat V}_{jj}({\vec v},\zeta ,\tau)
\right\}}\right],\,
\nonumber\\
{\hat V}_{jj}({\vec v},\zeta ,\tau)=
U\left[i\zeta_{j}(\tau)\hat{n}_{j}(\tau)-2\vec{v}_{j}(\tau)\cdot\vec{{\hat S}}_{j}(\tau)\right],\,
\qquad\qquad
\nonumber
\end{eqnarray}
here ${\vec v}_{j}$ and $\zeta_{j}$ are auxiliary fields connected with spin and charge on the site $j$. 

We consider the spiral type of incommensurate magnetic order, which is the rotation of order parameter in 
the plane, modulated with some wave vector ${\vec Q}=(Q_{x},Q_{y})$ (the superposition states of the rotation 
and the ferromagnetic component perpendicular to the plane have high energy \cite{Timirgazin_2009}). These 
states coincide with MF approximation states at zero temperature \cite{Igoshev_2010}
\begin{equation}
\label{eq:e_j}
{\vec e}_{j}={\vec e}_{x}\cos{({\vec Q}{\vec R}_{j})}+{\vec e}_{y}\sin{({\vec Q}{\vec R}_{j})}.\,
\end{equation}
The two-field  formalism allows us to obtain a Hartree-Fock approximation in the limit of  vanishing temperature 
\cite{Hassing_1973} and phase separation \cite{Igoshev_2010}. 
From here it is more convenient for us to make an identical transformation and to denote 
\begin{equation}
{\hat {\cal H}}_{0}=\sum_{{\vec k}}
{\cal E}_{{\vec k},\uparrow}\hat{c}_{{\vec k},\uparrow}^{+}\hat{c}_{{\vec k},\uparrow}^{}+
{\cal E}_{{\vec k},\downarrow}\hat{c}_{{\vec k},\downarrow}^{+}\hat{c}_{{\vec k},\downarrow}^{}
-UM\sum_{{\vec k}}
\left(\hat{c}_{{\vec k},\uparrow}^{+}\hat{c}_{{\vec k}+{\vec Q},\downarrow}^{}+
\hat{c}_{{\vec k}+{\vec Q},\downarrow}^{+}\hat{c}_{{\vec k},\uparrow}^{}\right),
\end{equation}
\vspace {-5mm}
\begin{equation}
{\hat {\cal H}}_{int}=U\sum_{j}\left[\left(\hat{n}_{j}-n_{e}\right)^{2}/4-
\left({\vec{\hat S}}_{j}\cdot\vec{e}_{j}-M\right)^{2}\right],\,
\nonumber
\end{equation}
with ${\cal E}_{{\vec k},\uparrow (\downarrow)}={\cal E}_{{\vec k}}^{}+Un_{e}/2$, 
${\cal E}_{{\vec k}}^{}=-2t(\cos{k_{x}}+\cos{k_{y}})$, where $n_{e}$ is a density and $M$ is a magnetization.
Introducing Green function for ${\hat G}_{0}(z)=(z-{\hat {\cal H}}_{0})^{-1}$ and making a standard calculation 
\cite{Wang_1969} we obtain 
\begin{eqnarray}
\label{eq:Z1}
Z^{*}[{\vec v},\zeta]=\exp{\left\{-\beta\Omega ({\vec v},\zeta )\right\}},\qquad
\Omega ({\vec v},\zeta )=\Omega_{0}-\frac{1}{\beta }
\sum_{n,j}Tr \ln{[1-G_{jj}^{0}(i\omega_{n}) V_{jj}({\vec v},\zeta)]}.
\end{eqnarray}
Here $\Omega_{0}=-1/\beta\ln{Tr\exp{[-\beta({\hat H}_{0}-\mu{\hat n})]}}$ is a thermodynamical potential 
of noninteracting electrons, $\omega_{n}=\pi(2n+1)/\beta$ are fermion Matsubara frequencies, 
$n=0,\pm 1,...$, $V_{jj}^{}$ and ${G}_{jj}^{0}$ are matrices  of spin variables. For simplification of the 
expression (\ref{eq:Z'}) we use the static approximation assuming that ${\vec v}_{i}(\tau)$ and 
$\zeta (\tau)$ are independent on $\tau$. We neglect the thermal fluctuation of the charge field 
$\zeta_{j}$. Thus, for each configuration of the spin field ${\vec v}_{j}$ we set $\zeta_{j}^{0}({\vec v}_{j})$ 
equal to the saddle point that is given by equation
\begin{eqnarray}
\label{eq:V0}
\qquad\qquad
\frac{\partial F({\vec v},\zeta)}{\partial\zeta_{j}}=U
\left[-2\zeta_{j}+in_{j}({\vec v_{j},\zeta_{j}})\right]=0,\qquad
\zeta_{j}^{0}({\vec v}_{j})=in_{j}({\vec v}_{j},\zeta_{j}^{0})/2,
\nonumber\\
n_{j}({\vec v}_{j},\zeta_{j})=\frac{1}{\beta}\sum_{n}Tr\left\{
\left[1-G_{jj}^{0}(i\omega_{n})V_{jj}({\vec v}_{},\zeta_{})
\right]^{-1}G_{jj}^{0}(i\omega_{n})\right\}.\qquad
\end{eqnarray}

We neglect longitudinal and leave only transverse fluctuations implying ${\vec v}_{j}=v\cdot{\vec e}_{j}$. 
We want to emphasize that ${\vec e}_{j}$ is the (arbitrarily chose)  $j$-dependent unit vector. At last, 
we change the matrix expression $(1- G_{jj}^{0}V_{jj})^{-1}G_{jj}^{0}$ for the average Green function 
${\tilde G}(z)$ (averaging is conducted over transverse fluctuations of auxiliary field $v$ $<...>_{v}$ ). 
The self-energy $\Sigma (z)$ and ${\tilde G}(z)$ is found from the self-consistency equation treatment by 
a coherent-potential approximation (CPA)
\begin{equation}
\label{eq:CPA}
{\tilde G}_{jj}(z)=<\left[1-G_{jj}^{0}(z)V_{jj}(v)\right]^{-1}
G_{jj}^{0}(z)>_{v}=G_{jj}^{0}(z-\Sigma(z)).
\end{equation}

At last we can write partition function in terms integration over $v$ 
\begin{equation}
\label{eq:Omega}
Z=\int d v\exp{\left\{-\beta U\left[v^{2}+\zeta_{}^{2}(v)\right]\right\}}
\exp{\left\{-\beta\Omega(v)\right\}},
\end{equation}
\vspace{-5mm}
\begin{eqnarray}
\Omega (v)=\Omega_{0}
+\frac{1}{\beta}\sum_{n} 
\left\{ln{\,det [1-{\tilde G}_{}(i\omega_{n})(V_{}(v) -\Sigma_{}(i\omega_{n}))]}-
ln{\,det [1+\Sigma_{}(i\omega_{n}){\tilde G}_{}(i\omega_{n})]}
\right\}.
\nonumber
\end{eqnarray}
For short of the expression (\ref{eq:Omega}) we miss out site index $j$.

Self-consistence solution of the Eqs.(\ref{eq:V0}),(\ref{eq:CPA}),(\ref{eq:Omega}) allows us to 
calculate  all magnetic properties of our system.
\begin{figure}[h]
\begin{center}
\includegraphics[scale=0.6]{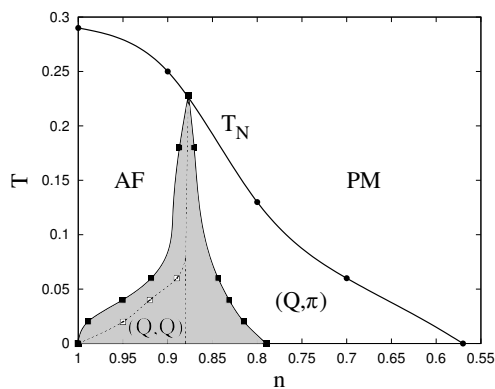}
\caption{Magnetic phase diagram for $U/t = 4$. Bold lines denote the temperature Neel (second-order phase 
transitions), solid lines denote boundaries of phase-separated region AF+ $(Q,\pi )$ (shaded area), 
dashed lines denote first-order phase transitions calculated without regard for PS.}
\end{center}
\end{figure}
\begin{figure}[h]
\begin{center}
\begin{minipage} [b]{0.45 \textwidth }
\includegraphics[scale=0.6]{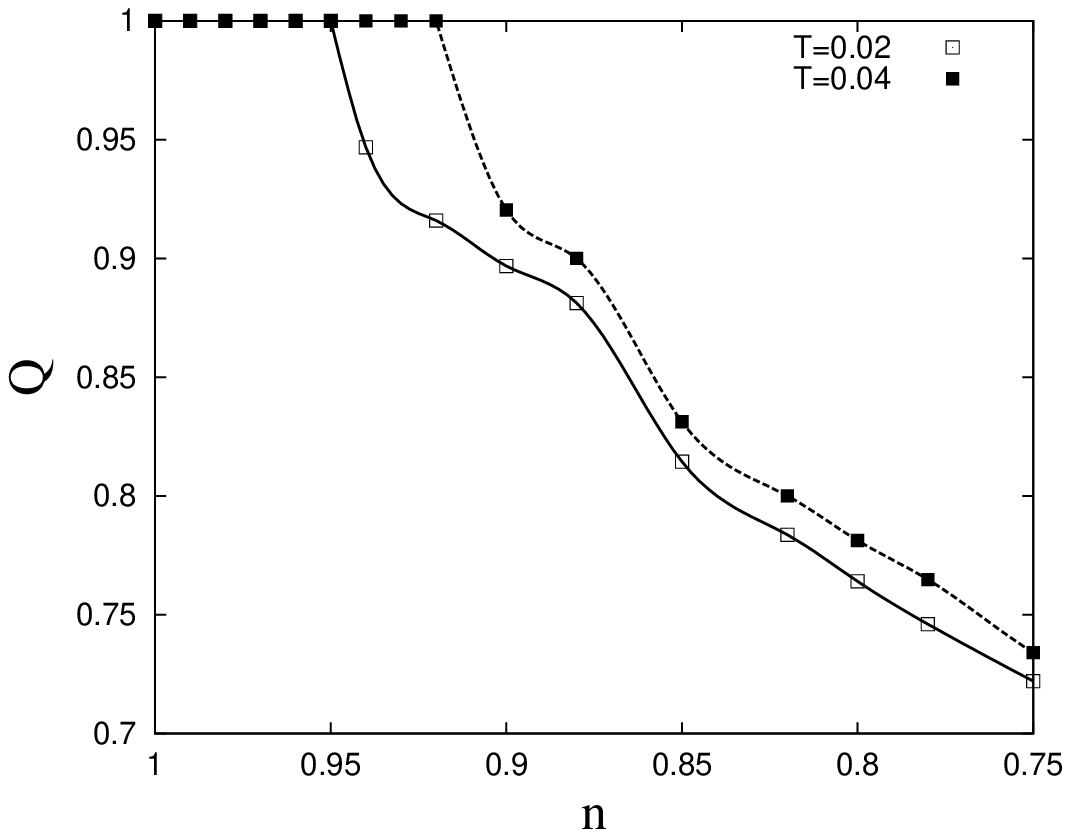}
\caption {}
\end{minipage}
\begin{minipage} [b]{0.45 \textwidth }
\includegraphics[scale=0.6]{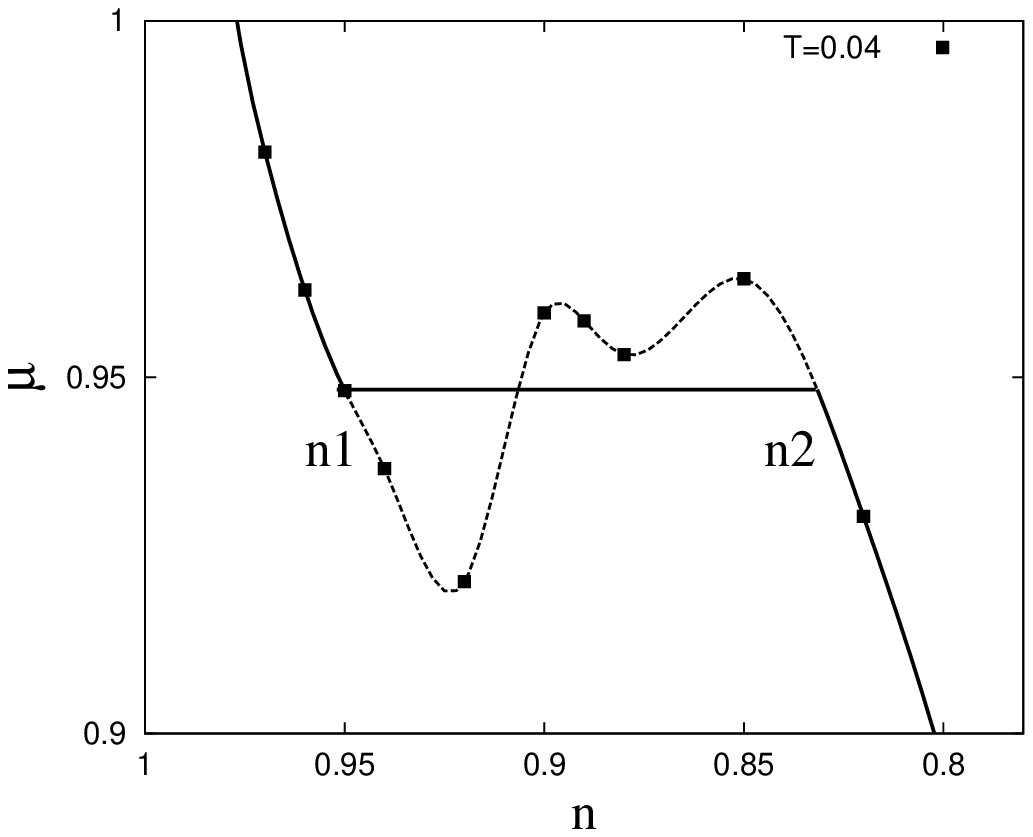}
\caption {}
\end{minipage}
\end{center}
\end{figure}

\section{Results}
\vspace {-3mm}

We have performed numerical calculations comparing thermodynamical potential of different magnetic 
phases at different $n$ and $T$, solving the Eqs. (\ref{eq:V0}),(\ref{eq:CPA}),(\ref{eq:Omega}) for 
$U/t=4$. Hamiltonian (\ref{eq:hamiltonian}) have the particle-hole symmetry $(n\leftrightarrow 2-n)$ 
and we can restrict ourselves to the region $0\le n\le 1$. The magnetic phase diagram is presented in 
Fig.~1. At zero temperature the antiferromagnetic (AF) phase exist only at half-filling $(n=1)$ 
\cite{Igoshev_2010,Timirgazin_2009}. The region of  AF states is intensively grown with temperature. 
The dependences of $Q_{x}$ on $T$ and n are shown in the Fig.~2. We can see that AF $\to (Q,Q)$ 
phase transition is a first-order transition. From our calculation it is clearly that $(Q,Q)\to (Q,\pi )$ 
and AF $\to (Q,\pi)$ transitions are also first-order transitions. This is confirmed by the dependence of chemical 
potential $\mu $ on number of electrons $n$ Fig.~3. We can see two regions of $\mu $  instability AF $\to (Q,Q)$ 
and $(Q,Q)\to (Q,\pi )$ phases, which are characterized by negative derivative of $\mu $ 
$ (n=(0.92-0.9, 0.87-0.86)) $. Therefore these regions must consist of two spatially separated 
phases. Boundaries of PS and proportion of phases can be obtained through the Maxwell construction. 
\begin{equation}
\label{eq:MAXWELL}
\int_{n1}^{n2}\left[\mu(n_{1})-\mu(n)\right] dn=0.\,
\end{equation}
The Fig.~3 shows dependence of the chemical potential with account to phase separation. 

\vspace{-3mm}
\section{Summary}
\vspace{-3mm}
\noindent In summary, our investigations demonstrate that the  magnetic phase transition 
qualitative coincides with experiments,  notably the transitions from paramagnetic state to AF 
one and then to PS at low temperature \cite{Yamada_1998,Matsuda_2000,Fujita_2002,Matsuda_2002}, 
see Fig.~1. However the phase separates to AF and $(Q,\pi )$ phases, whereas experiment shows 
separation to AF and $(Q,Q)$ phases and a region of separation to $(Q,Q)$ and $(Q,\pi )$ phases 
\cite{Matsuda_2002}. Behavior of the chemical potential (two regions of instability in Fig.~3) 
indicates a high possibility of the existence of PS AF~$+(Q,Q)$ and $(Q,Q)+(Q,\pi)$ phases at some 
choice of Hamiltonian parameters (\ref{eq:hamiltonian}), in particular with account of a 
next-nearest-neighbor hopping \cite{Igoshev_2010}. 

\vspace{-3mm}
\section{Acknowledgments}
\vspace{-3mm}
The work is supported No. 09-02-00461 from Russian Basic Research Foundation, and by Program  of 
RAS No.  09--2-2001


\vspace{-1.5mm}
\begin{thebibliography}{99}
\vspace{-2mm}
\bibitem{Yamada_1998} C.H.~Lee, K.~Kurahashi, J.~Wada, S.~Wakimoto, S.~Ueki, H.~Kimura, and 
Y.~Endoh: Phys. Rev. Vol. 57 (1998), p.~6165

\bibitem{Matsuda_2000} M.~Matsuda, M.~Fujita, K.~Yamada, R.J.~Birgeneau,
M.A.~Kastner, H.~Hiraka, Y.~Endoh, S.~Wakimoto, G.~Shirane: Phys. Rev. Vol. 62 (2000), p.~9148 

\bibitem{Fujita_2002} M.~Fujita and K.~Yamada, H.~Hiraka, P.M.~Gehring, S.H.~Lee, S.~Wakimoto, 
G.~Shirane: Phys. Rev. Vol. 65 (2002), p.~064505 

\bibitem{Matsuda_2002} M.~Matsuda, M.~Fujita, K.~Yamada, R.J.~Birgeneau, Y.~Endoh, G.~Shirane: 
Phys. Rev. B\, Vol. 65 (2002), p.~134515 

\bibitem{Igoshev_2010} P.A.~Igoshev, M.A.~Timirgazin, A.A.~Katanin, A.K.~Arzhnikov, and 
V.Yu.~ Irkhin: Phys. Rev. B\, Vol. 81 (2010), p.~094407 

\bibitem{Ghosh_1971} D.K.~Ghosh: Phys. Rev. Lett.\, Vol. 27 (1971), p.~1584 

\bibitem{Su_1996} G.~Su: Phys. Rev. B\, Vol. 54 (1996), p.~R8281 

\bibitem{Laad_2008} M.S.~Laad: arXiv0810.4416v1 (2008).

\bibitem{Langmann_2007} E.~Langmann and M.~Wallin:  Phys. Rev. Lett.\, Vol. 127 (2007), p.~825 

\bibitem{Macridin_2006} A.~Macridin, M.~Jarrell, and Th.~Maier: Phys. Rev. B\, Vol. 74 (2006), p.~085104 

\bibitem{Chang_2008} C.~Chang and S.~Zhang: Phys. Rev. B\, Vol. 78 (2008), p.~165101 

\bibitem{Maier_2005} T.~Maier, M.~Jarrell, T.~ Pruschke, M.H.~ Hettler, Rev. Mod. Phys,\, Vol. 77 (2005), 
p.~1027

\bibitem{Fasolino_2007} A.~Fasolino, J.H.~Los and M.I.~Katsnelson: Nature mat\, Vol. 78 (2007), p.~2011 

\bibitem{Timirgazin_2009} M.A.~Timirgazin and A.K.~Arzhnikov: Solid State Phenom. Vol. 559 (2009), 
p.~152-153  

\bibitem{Hassing_1973} R.F.~Hassing, D.M.~Esterling: Phys. Rev. B\, Vol. 7 (1973), p.~432 

\bibitem{Wang_1969} S.Q.~Wang, W.E.~Evenson, and J.R.~Schrieffer, Phys. Rev. Lett.\, Vol. 23 (1969), p.~92

\end{thebibliography}
\end{document}